\documentclass[11pt,twoside]{article}
\usepackage{asp2014}

\usepackage{eurosym}

\resetcounters

\begin{document}

\title{COTS software in science operations, is it worth it?}
\author{William  O'Mullane$^1$ and
Nana Bach$^2$ and Jose Hernandez$^1$ and Alexander Hutton $^2$ and Rosario Messineo$^3$ ,
\affil{$^1$  European Space Astronomy Centre,
P.O. Box 78,
28691 Villanueva de la Ca\~nada, Spain }
}
\affil{$^2$ Aurora for ESA at ESAC }

\affil{$^3$ Altec Turin, Italy }
\section{Introduction}

The Gaia astrometric satellite is now in operations for more than two years. The first data release \citep{2016arXiv160904172G} was highly successful. Behind the operations there is a lot of software and collaboration. 
The Data Processing Ground Segment is jointly operated by ESAC and the Gaia DPAC. This is comprised of the Science Operations Centre (SOC) operated by ESA and a set of Data Processing Centres (DPCs). The SOC is also a DPC (DPCE for daily and astrometric processing), SOC also  acts as the interface between the Mission Operations Centre (MOC) and the DPAC (See Figure \ref{socesac}). 
Gaia produces an impressive volume of raw data with about 50GB of uncompressed science data per day, yielding at mission completion a telemetry data volume of roughly 500TB. Transforming the data into scientifically meaningful quantities is the task of the Data Processing Analysis Consortium (DPAC). 
DPAC is comprised of a number of Coordination Units (CUs). Each one is responsible for a well-defined part of the Gaia data processing \citep{2008IAUS..248..224M}. 
For each CU there is at least one Data Processing Center (DPC) with dedicated resources for the data processing of the CU.

\articlefigure[width=.7\textwidth]{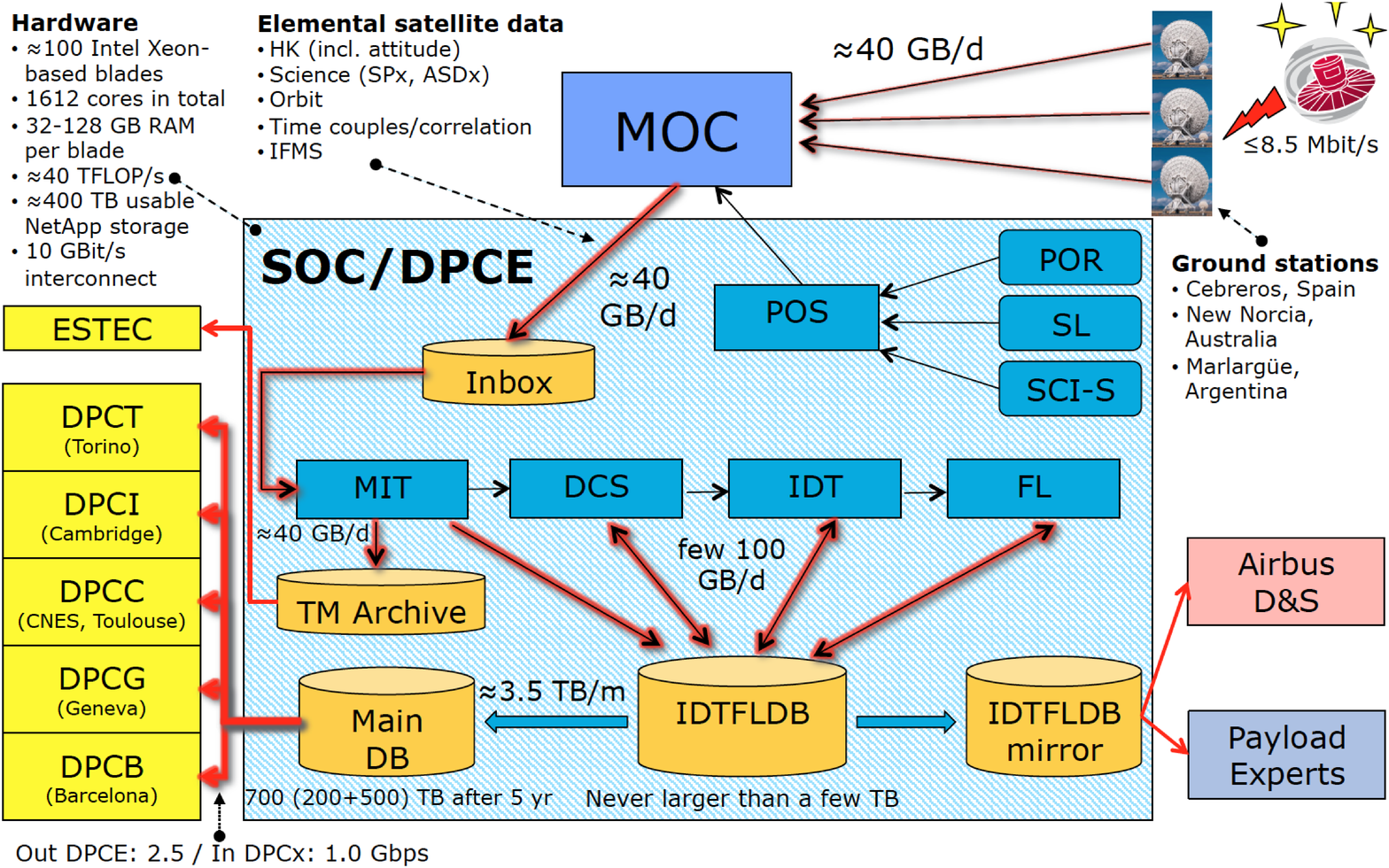}{socesac}{ Gaia SOC/DPCE at ESAC}

\section {The Main Database}

The MDB (Main Database) is the central repository of all the data produced by Gaia Scientific Processing and DPAC. The MDB Dictionary, running at DPCE, is used by all to define the Datamodel of the MDB.
There are actually a bunch of software systems incorporated in the MDB:
\begin{itemize}
\item MDB Database
\item MDB Schema Creator 
\item MDB Dictionary Tool 
\item MDB Extractor
\item MDB Ingestor
\item MDB Integrator
\item MDB Explorer 
\item MDB Data Manager

\end{itemize}
There are documents listing MDB Requirements and they may be summarised as follows:
\begin{itemize}
{\bf
\item 14 functional requirements 
\item 1 performance requirement
}
\begin{itemize}
\item MDB Schema Creator: 1 functional requirement
\item MDB Dictionary Tool: 24 functional requirements
\item MDB Extractor: 12 functional requirements, 1 performance requirement
\item MDB Ingestor: 10 functional requirements, 2 performance requirements 
\item MDB Integrator: 9 functional requirements
\item MDB Explorer: 16 functional requirements 
\item Organically added functionality
\begin{itemize}
\item MDB DataManager: no written requirements 
\end{itemize}
\end{itemize}
\end{itemize}

\section{Gaia Transfer System (GTS)}

GTS (Gaia Transfer System) is the service that permits data exchange between the Data Processing Centres (DPCs). This is  composed of Aspera (commercial)  and the DTSTool (add on built within DPAC by Altec). 
Aspera is a COTS for fast data transfer, it provides the technical platform for the data transfers. We have found this an extremely useful product with very good support. 

 The DTSTool is responsible for building an interface between Aspera and the data processing software systems at each DPC.

Again we have documents containing GTS Requirements which may be summarised as:
\begin{itemize}
{\bf
\item 13 functional requirements
\item 2 performance requirements
}
\end{itemize}

\begin{itemize}
\item 29 other requirements: 
\begin{itemize}
\item Portability: 1
\item User Interface: 5
\item Execution and activation: 1 
\item File naming convention: 3
\item Interface: 5
\item Safety and Security: 12
\item Performance: 3

\end{itemize}
\end{itemize}

\section{Cost-effectiveness analysis - factors}
For serious implementations, the real cost of software goes far beyond the license itself, and includes (see also \citet{P2_8_adassxxvi}):
\begin{itemize}
\item Time for implementation, evaluation, and integration, heavily influenced by:
	\begin{enumerate}
	\item The quality of documentation
	\item {\rm The responsiveness of support staff}
	\item The availability of code examples and other learning aides
	\end{enumerate}
\item Resources consumed by the use of the software.
\item Reliability of the software across all possible use-cases (often the most {\em expensive} aspect of cheaper software).
\item Flexibility and adaptability of the software relative to the competition.
\end{itemize}
To achieve real cost-effectiveness we recommend seeking  the following qualities in server software for high-demand workflows:
\begin{itemize}
	\item Memory-allocation independent (application terminates after execution).
	\item Fully separable from other functions.
        \item Availability of well defined and stable API in order to decouple the COTS from SW customisation.

	\item Proven track-record of reliability in high-volume production environments.
\end{itemize}

\subsection{MDB Cost}
We have a reasonable account of effort booked to the MDB work package  2005-2016:
	\begin{itemize}
	\item 	Effort $\approx 18 $ person years  incl. testing and documentation/support.
	\item Depending on our cost model that is a cost of \euro 2.8M to \euro 4.5M
Bear in mind this must include all consumables, office space, phones, travel etc.. It also contains management overhead not just an individuals salary.
	\item Lets call it \euro 3M.

	\end{itemize}

\subsection{GTS Cost}
For Aspera we have a mix of licenses and development:
\begin{itemize}
\item  Aspera licenses and support to date \euro 158K 
\item  DTSTool, customisation, automation etc. .. effort $\approx 3.5 $ person years incl. testing documentation.
\item Effort cost (using a similar rate to MDB but actually probably lower) \euro 560k 
\item Total cost \euro $\approx  720k $ 
\end{itemize}

\subsection{Cost effectiveness}

There is No good way to do this!  Constructive Cost Model (COCOMO) good {\em BEFORE} you build perhaps.
Though similar in requirement the MDB feels much more complex than GTS if we use the functional requirements and say all were met that gives MDB a much higher complexity with 86 v 29 requirements. 
\begin{itemize}
\item   UNIT COST = Cost/Function

\item MDB UC= 3M/86 = 34883

\item GTS UC= 720K/29  = 24827
\end{itemize}

COST EFFECTIVENESS  could be considered as the Ratio of in house to COTS software unit development cost. This would make COTS 1.4 times more cost effective.
Of course we can only use  COTS where appropriate, we  could not find a tool or set of tools to do the job of the  MDB.

\section{Conclusion}

To do this more effectively we should try harder to scope functional requirements in comparable way.
The COTS items we can use in space science is fairly limited (DBMS, xfer other generic stuff ).
But we should choose carefully - that small development can cost a lot cumulatively over our very long projects.

Open source is ok but can also die out over our long projects. In Gaia  we supported Apache Common Math - now ONLY we support it as the community seems to have disappeared.
	Java as a sort of open platform worked out ok for Gaia do far - who knows what will happen in the next decade.
	The Eclipse IDE , MANTIS (now we use Jira) are all good there are peripheral tools which work and save a lot.

\bibliography{O5-2.bib}

\end{document}